\documentclass[11pt,journal, onecolumn, draftcls]{IEEEtran}

\usepackage{hyperref}
\usepackage{amsmath}
\usepackage{amssymb}
\usepackage{graphicx}
\usepackage{cleveref}
\usepackage{cite}
\usepackage{color}
\usepackage{csquotes}

\usepackage{tikz}
\usetikzlibrary{shapes}
\usepackage{caption}
\usepackage{subcaption}
\newsavebox{\foobox}
\newcommand{\slantbox}[2][.5]
  {%
    \mbox
      {%
        \sbox{\foobox}{#2}%
        \hskip\wd\foobox
        \pdfsave
        \pdfsetmatrix{1 0 #1 1}%
        \llap{\usebox{\foobox}}%
        \pdfrestore
      }%
  }

\newcommand{\MDcomment}[1]{{\color{black}{#1}}}
\newcommand{\FS}[1]{{\color{black}{#1}}}
\newcommand{\JT}[1]{{\color{black}{#1}}}
\newcommand{\corr}[1]{{\color{black}{#1}}}

\crefname{equation}{}{}

\usepackage{macros}

\begin{document}

\title{Imaging with Equivariant Deep Learning}
\author{Dongdong~Chen, Mike~Davies,  Matthias~J.~Ehrhardt, Carola-Bibiane~Sch\"onlieb,~Ferdia~Sherry and~Juli\'an~Tachella
\thanks{D. Chen and M. Davies are with the School of Engineering, University of Edinburgh, Edinburgh, UK.}
\thanks{M. J. Ehrhardt is with the Department of Mathematical Sciences, University of Bath, Bath, UK.}
\thanks{C-B. Sch\"onlieb and F. Sherry are with the Department of Applied Mathematics and Theoretical Physics, University of Cambridge, Cambridge, UK.}
\thanks{J. Tachella is with the Centre National de Recherche Scientifique (CNRS) and \'Ecole Normale Sup\'eriere de Lyon, Lyon, France.}
}

\markboth{March 2022}%
{Shell \MakeLowercase{\textit{et al.}}: Bare Demo of IEEEtran.cls for IEEE Journals}

\maketitle

\begin{abstract}
From  early  image processing  to  modern  computational  imaging,  successful  models  and  algorithms have relied on a fundamental property of natural signals: \emph{symmetry}. Here symmetry refers to the invariance property of signal sets to transformations such as translation, rotation or scaling.  Symmetry  can  also be  incorporated  into  deep  neural networks in the form of equivariance, allowing for more data-efficient learning. While there has been important  advances  in  the  design  of  end-to-end  equivariant  networks  for  image  classification  in  recent years,  computational  imaging  introduces  unique  challenges  for  equivariant network  solutions  since  we typically  only  observe  the  image  through  some  noisy  ill-conditioned  forward  operator that  itself  may not  be  equivariant.  We  review  the  emerging  field  of  equivariant  imaging  and  show  how  it  can  provide improved generalization and new imaging opportunities. Along the way we show the interplay between the acquisition physics and group actions and links to iterative reconstruction, blind compressed sensing and self-supervised learning.
\end{abstract}


\IEEEpeerreviewmaketitle

\section{Introduction}

Traditional reconstruction methods as old as Wiener filtering exploit the symmetry principle by assuming that signals lie on a translation invariant subspace. Subsequent methods, such as wavelets~\cite{mallat1999wavelet}, go beyond translation invariance, accommodating other symmetries such as scale and rotations (e.g., steerable wavelets~\cite{simoncelli1995steerable}). Symmetry also plays an important role in image models based on partial differential equations~\cite{alvarez1993axioms}, where a basic list of invariance axioms for a multiscale analysis are shown to imply the existence of an underlying partial differential equation of a very particular form. \MDcomment{Note that in signal processing the terms invariance and equivariance are often conflated (the formal distinct mathematical definitions will be explained in Section~\ref{sec:group_actions} below). For example, the commonly used term linear time-invariant (LTI) systems should strictly be linear time-equivariant. In this setting, time equivariance has a simple meaning: inputting a time translated version of a signal, will output the same signal as the original but itself time translated.  }

More recent learning-based methods unveil another fundamental property of successful models and algorithms: \emph{learning from data}. Patch-based methods~\cite{wen2017frist} and convolutional sparse coding~\cite{wohlberg2016csc} eventually overtook wavelets, by relying both on symmetry --- invariance to translations and in some cases rotations --- and training data. These models can be seen as shallow predecessors of deep convolutional networks.

\begin{figure}
    \centering
    \includegraphics[width=1\linewidth]{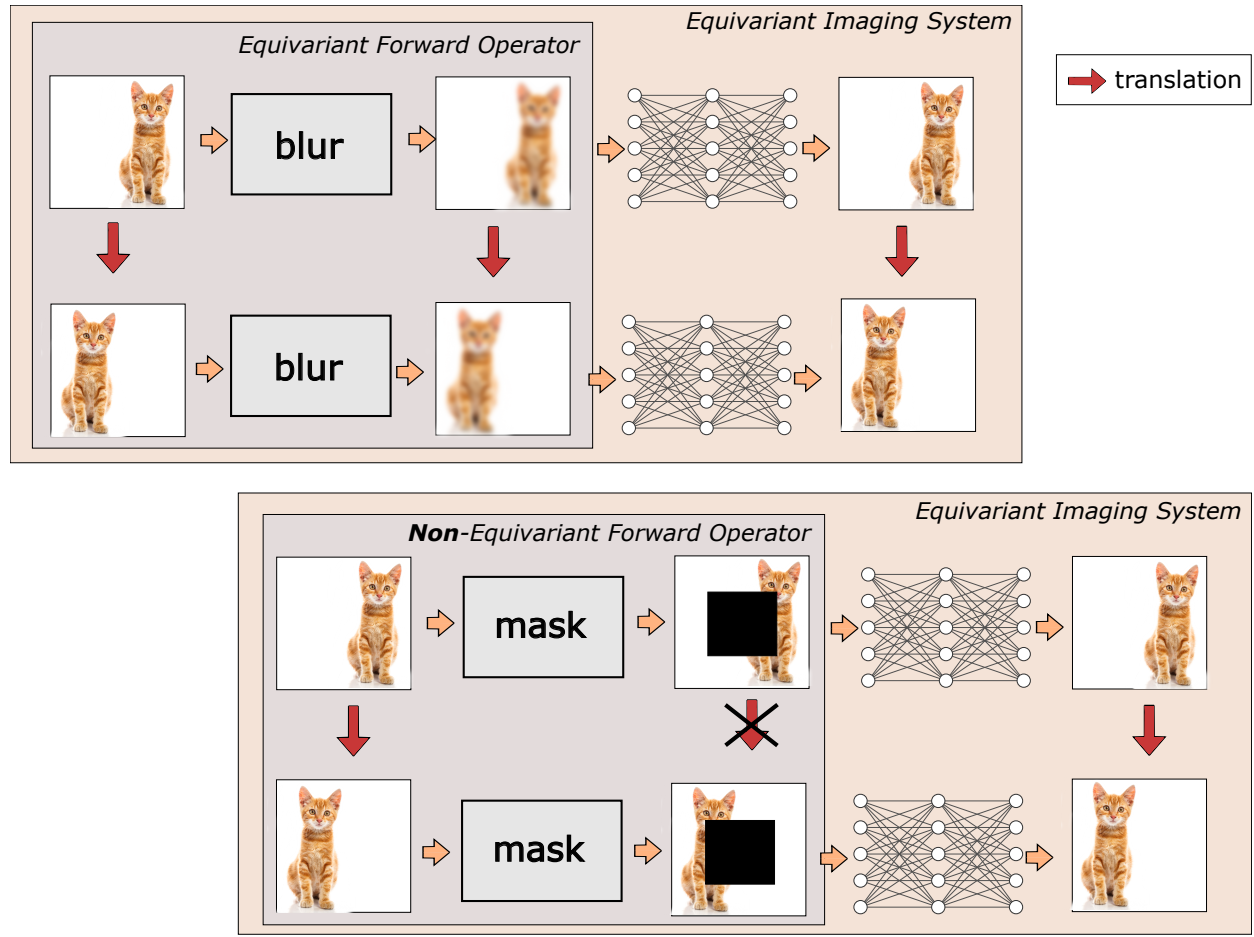}
    \caption{\textbf{The role of equivariance in computational imaging.} In imaging problems where the measurement process is equivariant to translations, such as the deblurring setting (top), a translation equivariant network (e.g., a standard CNN without pooling) generalizes to unseen translations, thus performing better than a non-equivariant counterpart (e.g., a fully connected network). However, in problems where the measurement process is not equivariant, such as the inpainting setting (bottom), a translation equivariant network \emph{would not} generalize to unseen translations, and a more careful analysis is needed. \corr{Specifically, a reconstruction network needs to achieve overall system equivariance, which differs from reconstruction equivariance when the forward operator is not itself equivariant (i.e., the inpainting case). This paper presents two approaches to achieve this: equivariant by design (\Cref{sec:equiv_design}) and equivariant by learning (\Cref{sec: equiv learning}).}
}
    \label{fig:equiv_summary}
\end{figure}

Deep neural networks (DNNs) are now ubiquitous in computational imaging and symmetry plays a fundamental role: almost all state-of-the-art networks include the powerful convolutional neural network (CNN) structure  which \JT{(without max pooling)} is equivariant to discrete translations by construction. For example DnCNN~\cite{zhang2017beyond}, a popular denoising network, is such a CNN.

The notion of equivariance generalizes to a much broader set of transformations beyond translations
\MDcomment{and can be exploited is many machine learning tasks. For example, AlexNet \cite{krizhevsky2012imagenet}, one of the early deep learning successes for image classification increased the ImageNet data by a factor of 2048 through a process of data augmentation (DA), generating new samples by applying flips, scaling and rotations to training images while leaving the labels unchanged. This has subsequently been shown to have learned representations within the network that are equivariant to these tranformations \cite{Lenc2015}.
} A network which is equivariant naturally generalizes to data with unobserved transformations, reducing sample complexity and outperforming non-equivariant counterparts~\cite{sannai2021improved}. In this paper, we present recent advances in building such equivariance into neural networks either by architectural design or through the training process. In particular, we review the design of equivariant networks~\cite{cohen2016group} which build invariances, such as rotations and flips, into the network architecture. Such networks can be directly used to solve computational imaging problems where the forward operator is also equivariant. For example, in the setting of deconvolution, a translation invariant reconstruction network generalizes to unseen translations of the training data.
However, in many computational imaging problems the measurement process is not itself equivariant, as illustrated in \Cref{fig:equiv_summary}. We therefore review strategies to overcome this and develop equivariant neural network solutions for general computational imaging problems. The first method, motivated by unrolled optimization algorithms, relies on building equivariant proximal subnetworks within the reconstruction network~\cite{celledoni2021equivariant}.

A second approach aims to enforce equivariance of the full imaging pipeline through training, either by  DA~\cite{fabian2021MRAugment} in the case of supervised learning, or through a carefully designed loss~\cite{chen2021ei} when tackling the unsupervised learning problem. The latter approach demonstrates that equivariance is a powerful strategy for fully unsupervised learning in computational imaging, as it only requires measurement data for training. This is extremely important as in many applications such as medical and scientific imaging, ground truth  data is either hard or impossible to obtain.

The paper is organized as follows: \Cref{sec:group_actions} introduces some basic mathematical concepts of group actions and invariance/equivariance. \Cref{sec: general CI} reviews the role of invariance/equivariance in computational imaging and discusses how deep learning is currently used to solve computational imaging problems. \Cref{sec:equiv_design} presents the main approaches to building equivariance into a network's architecture and \Cref{sec: equiv learning} shows how to enforce equivariance during training and its implications for unsupervised learning. Open problems are discussed in \Cref{sec: open problems}. 

\section{Group Actions and Equivariance}\label{sec:group_actions}
We will use this section to explore some of the basic concepts that are needed to give a proper treatment of this topic. The concept of symmetry is usually mathematically formalised through the definition of an algebraic object called a group. A group $(G, \cdot)$ is a set $G$ equipped with a product $\cdot : G\times G \to G$ that is associative, $ g_1\cdot(g_2\cdot g_3) = (g_1\cdot g_2)\cdot g_3$, with the additional requirements that there is an identity element $e\in G$ satisfying $g\cdot e = e \cdot g = g$, and that for each element $g\in G$ there is an inverse element $g^{-1}\in G$ such that $g\cdot g^{-1} = g^{-1} \cdot g = e$. When there is no risk of confusion, we may drop the dot for the group product and simply write $g\cdot h = gh$, and we will refer to the group by the name of the underlying set $G$.

The concept of a group is particularly interesting when combined with the concept of an action: given a (potentially abstract) group $G$ and a set $X$, we will say that $G$ acts on $X$ through $T$ if $T = \{T_g: X\to X \}_{g\in G}$ is a collection of invertible transformations that is compatible with the group, in the sense that $T_{g_1}\circ T_{g_2} = T_{g_1\cdot g_2}$. That is, a group action turns an abstract group into a group that can be identified as a concrete set of transformations. Any given group may act on many different sets $X$, and may even act on the same set in many different ways. A particularly simple group action is the \emph{trivial} action of $G$ on $X$, in which the case the group ``acts'' by doing nothing: $T_g = \id_X$ for all $g\in G$.

\begin{figure}[ht!]
 \newcommand{\transpic}[4]{
  \tikz\node [rectangle, draw, black, very thick, minimum size=3.3cm,
    path picture = {
      \node [#1] at (path picture bounding box.center) {
        \slantbox[#2]{\includegraphics[#3]{#4.png}}};
    }] {};}

    \centering

    \begin{subfigure}[b]{0.3\textwidth}
    \centering
    \transpic{}{0}{width=2.3cm}{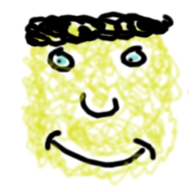}
    \caption*{reference}
    \end{subfigure}
    \begin{subfigure}[b]{0.5\textwidth}
    \centering
    \transpic{fill=black}{0}{width=2.3cm}{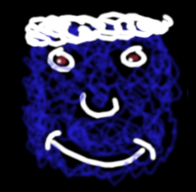}
    \caption*{transformation of range: e.g. color inversion}
    \end{subfigure}\\[4mm]%

    \begin{subfigure}[b]{\textwidth}
    \centering
    \transpic{xshift=0.6cm,yshift=0.5cm}{0}{width=2.3cm}{fig2a}\hspace{1mm}%
    \transpic{}{0}{width=2.3cm,angle=-45}{fig2a}\hspace{1mm}%
    \transpic{}{0}{width=3.5cm}{fig2a}\hspace{1mm}%
    \transpic{xshift=1cm}{-.7}{width=2.3cm}{fig2a}
\caption*{transformation of domain: e.g. translation, rotation, scaling, shearing}
    \end{subfigure}%
    \caption{\textbf{Illustration of group actions on images.} Groups can act on images either by transforming their range or their domain.}
\label{fig:groupactions}
\end{figure}

In this work, we are concerned with images, in which case the signals of interest can usually be modeled as functions $u:X\to Y$. As an example, for color images $X$ is a subset of $\eR^2$ and $Y$ is $\eR^3$. A group action $T$ of $G$ on the domain $X$ can be lifted to an action $T'$ on the set of signals by $T_g'(u)(x) = u(T_{g^{-1}}(x))$. If the set of signals is a vector space (as for the color images), $T'$ acts linearly on the signals, making it a so-called representation of $G$. Similarly, an action $T$ of $G$ on the range $Y$ of the signals can be lifted to an action on the signals, $T'$, by $T'_g(u)(x) = T_g(u(x))$, and in fact actions on the domain and range can be combined if so desired. For some examples of group actions lifted from the domain $X = \eR^2$ or range $Y=\eR^3$ to signals $u:\eR^2\to\eR^3$, the reader is referred to Figure~\ref{fig:groupactions}. All of the domain transformations shown in the figure are examples of affine transformations; the affine group on a Euclidean space $\eR^d$ consists of all transformations of the form $x\mapsto Hx + h$, where $H\in \eR^{d\times d}$ is an invertible matrix and $h\in \eR^d$ is a translation, with the group product given by composition. Important subgroups of the affine group include the group of translations, where $H=\id_{\eR^d}$, and the group of roto-translations, where $H$ is restricted to be a rotation matrix. When a group $G$ acts linearly on a set $X$ through an action $T$ (as in the cases highlighted above), we may drop parentheses and simply write $T_gx$ to mean $T_g(x)$. \corr{Moreover, if $G$ is compact, under an appropriate basis, the action is orthogonal, such that  $T_{g^{-1}}=T_g^{-1}=T_g^{\top}$.}

In the setting of computational imaging, we are concerned with maps $\Phi: \mathcal X\to \mathcal Y$ between potentially different sets $\mathcal X$ and $\mathcal Y$ representing spaces of images and/or measurements. For example, $\Phi$ could be the forward operator or a reconstruction operator. If both $\mathcal X$ and $\mathcal Y$ share a symmetry in the form of potentially different group actions $T$ of $G$ on $\mathcal X$ and $T'$ of $G$ on $\mathcal Y$, we may ask whether $\Phi$ respects these symmetries, in the following sense:

\textbf{Equivariance:} we call $\Phi$ equivariant, if $\Phi(T_g(u)) = T'_g(\Phi(u))$ holds for all $u\in \mathcal X$ and $g\in G$.

\textbf{Invariance:} if $T'$ is the trivial action of $G$ on $\mathcal Y$ and $\Phi$ is equivariant, we will call $\Phi$ invariant. In this case, we have $\Phi(T_g(u)) = \Phi(u)$ for all $u\in \mathcal X$ and $g\in G$.

As an example, that we will elaborate on further in Section~\ref{sec:equiv_design}, it is natural to require that a denoiser $\Phi:\mathcal X\to \mathcal X$ satisfies a group equivariance property when the dataset of images that we are considering has a group invariance property.

An additional fact that will be of importance later is that equivariance is preserved under function composition: if $G$ is a group that acts on spaces $\mathcal X, \mathcal Y, \mathcal Z$ through $T, T', T''$ respectively and $\Phi: \mathcal X\to \mathcal Y$ and $\Psi: \mathcal Y\to \mathcal Z$ are equivariant, $\Psi \circ \Phi : \mathcal X\to \mathcal Z$ is equivariant in the sense that $(\Psi\circ\Phi)(T_g(u)) = T''_g((\Psi \circ \Phi)(u))$.

\section{Computational Imaging, Equivariance and Deep Learning}\label{sec: general CI}

Computational imaging, distinct from other forms of image processing, relies on the acquisition of sensor measurements that indirectly inform about the imaged object. Reconstruction therefore requires some form of inversion of the physical acquisition process via an imaging algorithm. Computational imaging systems span a broad range of applications, such as computational microscopy, medical imaging (CT, MRI, ultrasound imaging), computational photography, synthetic aperture radar, geophysical imaging and seismic imaging.

Such imaging systems aim to solve a mathematical \emph{inverse problem} to reconstruct the (continuous) image, $u$, from a discrete number of measurements, $y\in\eR^m$. In order to facilitate this computation it is necessary to represent the image in discrete form, e.g., through an appropriate basis function expansion, therefore with a slight abuse of notation that should be clear from the context, we will also describe the image representation as the finite dimensional vector $u \in \eR^n$ that can be estimated through a stable \enquote{inversion} of the forward (acquisition) process $A$:
\begin{equation}\label{eqs:forward_model}
    y = A(u) + \epsilon.
\end{equation}
where $\epsilon$ captures any noise or modelling errors. Although early computational imaging systems utilized linear or analytical reconstruction algorithms such as filtered back projection or the nonuniform FFT, modern imaging systems have taken advantage of more sophisticated reconstruction algorithms that enable imaging with subsampled and noisy sensor measurements \cite{ravishankar2019image}.

The task is particularly challenging when there is substantial noise and when the forward operator $A$ is ill-conditioned or rank deficient (e.g., when $m< n$). Here the forward operator $A$ models all the physics of the acquisition process, e.g., in MRI this may include the influence of the excitation pulse sequence, the form of k-space sampling, and any uncertainties such as the coil sensitivities. In this tutorial, we will mostly consider a linear $A$ although many ideas presented extend to more general operators.

\MDcomment{\subsection{Model Based Image Reconstruction}\label{sec: DL for CI}

A popular appproach to tackling such challenging inverse problems has been to adopt a model-based image reconstruction (MBIR) methodology by solving a regularized variational optimization problem that is composed of a data consistency loss term $E(u) = d(A(u),y)$ to capture the role of the acquisition physics in the measurement process, including the noise statistics,} along with a regularization function $J(u)$ that incorporates prior knowledge (e.g., sparsity in the wavelet domain) of $u$ and penalizes less plausible solutions. MBIR therefore typically aims to solve optimization problems of the following form:
\begin{equation}\label{eqs:rls}
    \argmin_u E(u) +  J(u).
\end{equation}

While much attention has been paid during the last two decades to the important role of sparsity and related  low-dimensional models for $u$, due to their ability to solve such ill-posed inverse problems, it is important not to lose sight of the geometric nature of the underlying models, \MDcomment{which almost invariably encodes aspects of the image physics through structure and symmetry.} For example, in the continuous domain, the popular total variation (TV) prior~\cite{rudin1992} is naturally invariant to translations, rotations, and reflections. Similarly, wavelet models capture translation and scale invariance. Less obviously, popular patch based models, e.g.~\cite{ravishankar2011,wen2017frist}, incorporate powerful patch permutation invariance, as well as the more obvious discrete translation invariance.

MBIR methods usually have to be solved iteratively, and for scalability usually focus on first-order optimization strategies. \MDcomment{While the data consistency loss is generally a smooth function, state-of-the-art regularizers, such as those that enforce sparsity or low-rank solutions are non-differentiable and therefore simple gradient descent methods cannot be directly applied. A popular set of solutions for such problems that we will make use of later are proximal splitting methods \cite{Combettes2011}. In such methods the objective function is split into multiple terms that can be handled separately through easier subproblems.}

For example, a non-differentiable regularization function $J$ can be handled through its proximal mapping which takes the following form:
\begin{equation}\label{eq: prox}
\mbox{prox}_J (u) = \argmin_v \frac{1}{2} \|u-v\|_2^2 +J(v)
\end{equation}
Intuitively this can be seen as a generalization of the projection operator~\cite{Combettes2011}. Various splitting methods have now been developed and include proximal gradient descent (PGD) and its fast variants, alternating direction method of multipliers (ADMM), and numerous others \cite{Combettes2011}. \MDcomment{For example, the basic PGD algorithm proceeds by taking a step in the negative gradient direction of the smooth component of the cost function, followed by a proximal mapping to reduce the non-differentiable cost. Its update equation thus takes the following simple form:}
\begin{equation}\label{eq:pgd}
    u^{(k)} = \mbox{prox}_{\tau J} \left(u^{(k-1)}-\tau \nabla E(u^{(k-1)})\right)
\end{equation}
where $\tau > 0$ denotes the step size of the algorithm.

\subsection{Deep Learning for Inverse Problems}\label{sec: DL for CI}

While advanced MBIR algorithms have demonstrated impressive results, particularly for ill-posed inverse problems, they are limited by our ability to construct effective prior models, and also by the computational complexity of solving the subsequent iterative optimization solutions (despite huge improvements in recent years) and hyper-parameter selection. This has in part driven researchers to explore alternative data-driven reconstruction methods based around machine learning.

In particular, due to the powerful representation learning properties of DNNs, a range of neural network solutions have recently been proposed for computational imaging (see \cite{ongie2020deep, Arridge2019} for detailed surveys).
In this setting, the goal is usually to learn a reconstruction function $f_\theta: y\mapsto u$ parameterized by the network weights $\theta$, using $N$ pairs of measurements and ground-truth images $\{(y_i,u_i)\}_{i=1,\dots,N}$. The networks are typically trained by minimising the empirical risk
\MDcomment{
\begin{equation} \label{eq:std loss}
    \min_\theta \sum_{i=1}^{N} \ell(u_i,f_\theta(y_i))
\end{equation}
where a popular choice is the squared error loss $\ell(u_i, f_\theta(y_i)) = \|u_i-f_\theta(y_i) \|^2_2$, although other losses ($\ell_1$, perceptual, etc.) can also be used for training. }
Conceptually, the simplest approach is to use a neural network to directly predict the output image from the measurements: $\hat{u} = f_\theta (y)$. \MDcomment{In practice, it is common to nominally incorporate the acquisition physics by first mapping the measurements into the image domain, e.g., using the pseudo-inverse or another simple inverse operator when $A$ is linear, e.g.~\cite{jin2017deep}.} Such networks can then be trained without further exploiting the knowledge of $A$ in either training or testing. The general principle is that, given enough  data, we should be able to train a neural network to learn the mapping between $A^{\dagger} y$ and $u$ directly. While the success of this approach is likely to depend on the complexity and structure of the acquisition physics, it has been observed to work well for numerous tasks, such as CT imaging, superresolution and deblurring. However, this approach may require large quantities of training data because it is required to not only learn the geometry of the image space containing $u$, but also aspects of the operator $A$.

\MDcomment{Alternatively, we can design the architecture of $f_\theta$ using ideas of MBIR solutions (see the various review articles~\cite{ongie2020deep, ravishankar2019image, monga2021algorithm} for a complete list of such techniques).
For example a pre-trained generative network can be used to directly model the image prior, e.g.~\cite{bora2017generative}, and thereby incorporated into an MBIR algorithm. Alternatively, in plug-and-play reconstruction, based on interpreting the proximal operator as a signal denoiser, we can replace the proximal operator in the MBIR algorithm with a sophisticated data-driven denoiser, such as a pre-trained \JT{CNN}, e.g.~\cite{ahmad2020plug}. More generally, there has been a growing trend in building \JT{deep networks} by \emph{unrolling} a finite number of iterations of an MBIR algorithm and replacing various components by neural network computations. The weights of the resulting unrolled network can then be trained in an end-to-end manner using back propagation. While a number of variations have been proposed in the literature~\cite{monga2021algorithm}, here we will focus on the use of neural networks to replace the proximal operator associated with $J(u)$. This choice provides a natural separation of the algorithmic components that promote consistency with the measurements and that can easily exploit the known acquisition physics from those involving the image model which is less well defined and benefits far more from a data-driven approach. }

\begin{figure}[ht!]
    \centering
    \includegraphics[width=\linewidth]{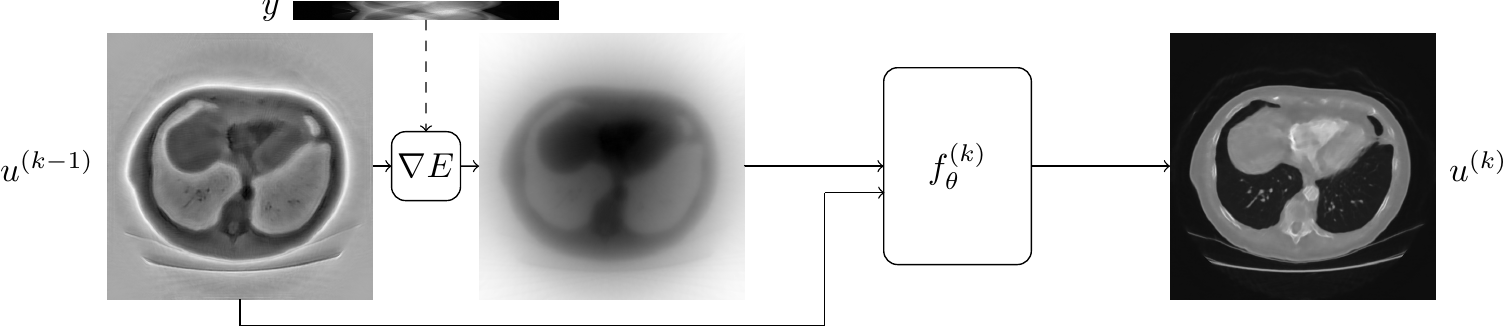}
    \caption{A schematic overview of an iteration of an unrolled PGD algorithm applied to the problem of computed tomography (CT) reconstruction. \corr{In this setting, the inputs $u^{(k-1)}$ and $\nabla E(u^{(k - 1)})$ may be combined as in PGD using a (learnable) step size $\tau$, to give $u^{(k-1)} - \tau \nabla E(u^{(k - 1)})$ before processing with the neural network. It is also possible, however, to allow the neural network to learn a more general mixing of these inputs.}
    }
    \label{fig:schematic_prox_grad}
\end{figure}

Consider again the PGD algorithm in~\Cref{eq:pgd}. A simple modification replaces the proximal map \corr{along with the step size, $\tau$,} at the $k$th iteration with a neural network $f_\theta^{(k)}$, such that:
\begin{equation}\label{eq:lpgd}
    u^{(k)} = f_\theta^{(k)} (u^{(k-1)},\nabla E(u^{(k-1)})).
\end{equation}
\corr{The algorithm is then run} for a fixed number of iterations, $k = 1,\ldots , \text{iter}_{\max}$, as illustrated in \Cref{fig:schematic_prox_grad}. The learnable weights in individual networks $\{f_\theta^{(k)}\}$ can be either tied or varied from iteration to iteration.

While such \MDcomment{hybrid MBIR-DNN} approaches have proved highly successful, providing state-of-the-art imaging solutions, it appears that by adopting these data-driven approaches we may have thrown away the other prior physical knowledge we have, namely the symmetry properties of our signal set. In the next two sections, we review ways to remedy this through either a modified unrolled network architecture or through the training process itself.

\section{Equivariance by Design}\label{sec:equiv_design}
Let us consider the problem of designing neural networks that are equivariant, building on the observation made in Section~\ref{sec:group_actions} that equivariance plays well with function composition: neural networks are alternating compositions of simple linear and nonlinear functions, so we are led to study the problem of designing linear and nonlinear equivariant functions. For clarity of exposition, this section will treat signals as continuous objects, although, as noted in \Cref{sec:group_actions}, we always deal with discrete data in practice and depending on the symmetry under consideration,  equivariance cannot be expected to always hold exactly after discretization.

One established approach to designing equivariant networks, which leads into the systematic approach that we will study in the next section, can be found in CNNs~\cite{lecun_convolutional_1998}. Treating an image as a function $u:\eR^2 \to \eR$, we can act on it with a translation $h\in \eR^2$ by $T_hu(x) = u(x - h)$. In this case, convolution by an arbitrary filter $k:\eR^2\to \eR$, i.e.\ \corr{$u \mapsto k*u$}, is equivariant, where
\corr{
\begin{align}
    k*u(x) = \int\limits_{\eR^2} k(x')u(x - x') \,\mathrm d x'.
    \label{eq:conv}
\end{align}}
\FS{Note that in practice, most deep learning software libraries expose a ``convolution'' operation that actually computes a cross-correlation operation.} In addition, any activation function applied pointwise is also equivariant, so a translation-equivariant neural network can be designed by alternating convolutions with (learnable) filter banks and pointwise application of activation functions. In practice, equivariance may be broken by incorporating additional operations such as downsampling and upsampling, although this issue can be overcome by using suitable replacements of these operations. This can be done for instance using adaptive polyphase upsampling and downsampling, and was applied to some Fourier-based computational imaging tasks in~\cite{Chaman2021}. \FS{It is worth noting that even so, edge effects that arise as a result of bounded image domains will always prevent exact translational equivariance from holding.}

\subsection{Equivariant Neural Networks}

As previously mentioned, systematic approaches to building group equivariance into neural networks can be found by separately designing equivariant linear maps and equivariant nonlinearities and composing them. Let us first consider the problem of designing equivariant linear maps. Broadly speaking, approaches to solving this problem can be split in two tracks, both of which build on the concept of convolution that has been used in ``ordinary'' CNNs previously.

\textbf{Lifting approach:} \FS{It is possible to generalise the Euclidean convolution of Equation~\eqref{eq:conv} to a group convolution, which combines two signals defined on the group in an equivariant manner. }Under a technical condition (local compactness) that is satisfied for many groups, it is possible to define an invariant measure $\mu$ (the so-called Haar measure) on the group $G$. This invariance means that for any integrable $u:G\to \eR$ and group element $g\in G$, we have
\begin{align}
    \int\limits_G u(gh)\,\mathrm{d} \mu(h) = \int\limits_G u(h) \,\mathrm d \mu(h).
\end{align}

\FS{Essentially the Haar measure should be thought of as a uniform measure on the group; if $G=\eR^d$ it coincides with the Lebesgue measure, whereas if $G$ is discrete it is simply the counting measure.} With this measure, we can define equivariant convolutions on the group by
\corr{
\begin{align}
    k*u(g) = \int\limits_G k(h)u(h^{-1}g) \,\mathrm d \mu(h).
    \label{eq:groupconv}
\end{align}
}
As with Euclidean convolutions, we can discretize such convolutions and parametrize the convolution kernel $k$ with learnable parameters. \FS{As a simple example, it is worth remarking that the Euclidean convolution in Equation~\eqref{eq:conv} is a special case of the group convolution in Equation~\eqref{eq:groupconv} when $G=\eR^2$ is the group of translations.}

Note that this convolution acts on signals that have as domain $G$. This is where the lifting name comes into play: an input signal such as an image generally has as domain a space such as $\eR^d$\FS{, i.e.\ it is not of the form required to apply the Equation~\eqref{eq:groupconv}. To prepare such an ordinary input signal, we need to ``lift'' it to $G$, for instance using a linear map such as}
\begin{align}
    Lu(g) = \int\limits_{\eR^d}  k(g^{-1}x)u(x)\,\mathrm dx,
\end{align}
where $k:\eR^d\to \eR$ is again a filter with learnable parameters. This approach was pioneered in~\cite{cohen2016group}, where it was applied to learn invariant image classifiers. In addition to the problem of lifting an input signal to the group, there is the opposite problem of projecting a signal on the group back to a signal on the original domain. This problem is of particular interest in computational imaging tasks in which we are designing image-to-image maps. Although it is possible to overcome this issue, there is another approach to designing equivariant linear maps that neatly bypasses it completely as we discuss next.

\begin{figure}
    \centering
    \includegraphics{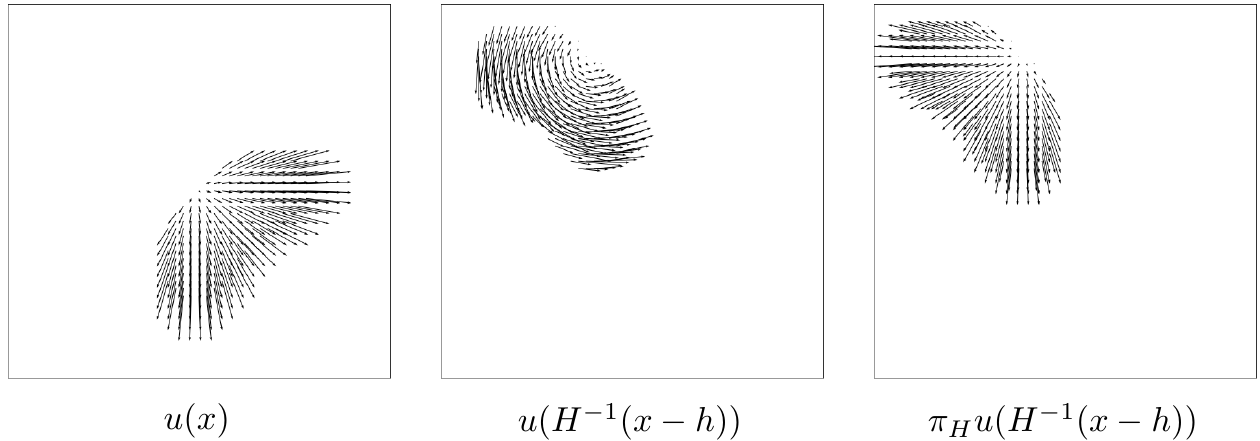}
    \caption{\FS{To properly transform geometric features, such as the vector field shown here, it is necessary for the group action on the domain to be followed by a group action on the range. In this case, we have a vector field, so the representation $\pi$ is simply given by $\pi_H = H$.}}
    \label{fig:vector_fields}
\end{figure}

\textbf{Steerable filters approach:} If the symmetry that we are interested in preserving is a subgroup of the affine group (recall its definition from Section~\ref{sec:group_actions}), we can use ordinary convolutions with kernels that are appropriately constrained to get equivariant linear maps. \FS{More specifically, we assume that this subgroup contains all translations}\corr{ and that the transformations are isometries}. In this setting, it is natural to consider signals that are not just scalar images, but also signals that represent higher-order geometric features. This is formalised by postulating that the group that we consider acts not only on the domain of the signal, but also on the range of the signal. As a simple example that illustrates this concept, let us consider a feature map of edge normal vectors of an image and what should happen to it under a roto-translation of the underlying image: not only should the domain be transformed, but the edge normals should be appropriately rotated as well. \FS{The necessity of this transformation of the range of a geometric feature is illustrated in more detail in Figure~\ref{fig:vector_fields}. For more complicated geometric features that are not just vector fields, this is generalised by transforming the range of the signal using a representation $\pi$ of the linear transformations being considered.} Correspondingly, the group actions we consider will all be of the form
\begin{align}
    T^{\pi}_{(h, H)}u(x) = \pi_H u(H^{-1} (x - h)),
\end{align}
where $(h, H)$ is a group element, consisting of a translation $h\in \eR^d$ and a linear operator $H: \eR^d\to \eR^d$, and \corr{$\pi_H$ is a representation of the linear operator $H$, acting} on the range of the signal $u:\eR^d \to \eR^{d_\pi}$.

\FS{We assume that there is such a group action on the input signals and a similarly defined group action $T^{\pi'}$ on the output signals for a potentially different representation $\pi'$ of the linear operators under consideration. The goal is to design an equivariant convolution mapping input signals $u_\text{in}:\eR^d \to \eR^{d_{\pi}}$ to output signals $u_\text{out}:\eR^d \to \eR^{d_{\pi'}}$ by}
\corr{
\begin{align*}
    k *u(x) = \int\limits_{\eR^d} k(x') u(x - x')\,\mathrm dx'
\end{align*}
Writing out the equivariance condition, we find that
\begin{align*}
    \pi_H' \int\limits_{\eR^d}k(x') u(H^{-1} (x - h) - x')\,\mathrm dx' &= T_{(h, H)}^{\pi'}[u * k](x)\\
    &=[T_{(h, H)} ^\pi u]*k(x)\tag{Equivariance}\\
    &= \int\limits_{\eR^d} k(x') \pi_H u(H^{-1}(x - h - x'))\,\mathrm dx'\\
    &= \int\limits_{\eR^d} k(Hx') \pi_H u(H^{-1}(x - h) - x')\,\mathrm dx' \tag{Change of variables, using that $H$ is an isometry}.
\end{align*}
Rearranging, we have
\begin{align*}
0 = \int\limits_{\eR^d} (\pi_H' k(x') - k(Hx') \pi_H) u(H^{-1}(x-h) - x')\,\mathrm dx'
\end{align*}
and since this must hold for arbitrary signals $u$,} we find that this is equivalent to asking that the kernel $k:\eR^d\to \eR^{d_{\pi'}\times d_{\pi}}$ satisfies the condition
\FS{
\begin{align}
    k(Hx) \pi_H = \pi'_H k(x)
\end{align}
}
for all linear operators $H:\eR^d\to \eR^d$ that occur in the group $G$~\cite{weiler_general_2019}. This constraint (which is linear in $k$) can be solved ahead of time, and discretized to give a basis of equivariant convolution kernels. Notably, when $d = 2$ and the group in consideration is a group of roto-translations, the constraint is equivalent to requiring that the kernel be decomposable into the product of a radial function and specific circular harmonics (essentially the same technique that was used in the original steerable wavelets~\cite{simoncelli1995steerable}). This is further discussed in~\cite{weiler_general_2019} and the associated software package implements these equivariant convolutions in a way that allows for easy experimentation.

Now that we have an idea of how equivariant linear maps can be designed, we will move on to the problem of equivariant nonlinearities. In many DNN architectures, nonlinearities take a particularly simple form: a scalar nonlinearity is applied component-by-component to the output vectors of linear maps. This approach can be used without problems to get equivariant nonlinearities too, as long as the group actions only act on the domain space of the signals. When the group acts nontrivially on the range space  of  the  signals as well  (as  in  the  steerable  filter  approach  above) it is necessary to ensure that  the nonlinearity is equivariant to the group action. This restricts the range of admissible nonlinearities, though a variety of solutions have been identified. \FS{An example of an equivariant nonlinearity between features of the same type (so that $\pi = \pi'$) is the norm nonlinearity: if $\pi$ is a unitary representation, meaning that $\|\pi_H v\| = \|v\|$, the norm nonlinearity $f$ maps an input signal $u:\eR^d\to\eR^{d_\pi}$ to an output signal $f(u):\eR^d\to \eR^{d_\pi}$ with $f(u)(x) = \phi(\|u(x)\|) u(x)$, with $\phi :\eR\to \eR$ a scalar nonlinearity.} We direct the interested reader to~\cite{weiler_general_2019} for more details on equivariant nonlinearities.

\subsection{Applications to Computational Imaging}

Many previous applications of equivariant neural networks have focused on such tasks as image classification and image segmentation. In these tasks, one of the main benefits observed from enforcing equivariance is a reduced sample complexity. Especially when training data is hard to get, the additional inductive bias from building symmetries into a machine learning method enables it to make much more efficient use of available data~\cite{sannai2021improved}. 

\begin{figure}[ht!]%
    \centering%
    \hspace*{-2mm}%
    \begin{tikzpicture}%
    \node at (0,0) {\includegraphics[width=.75\linewidth, clip, trim=0pt 0pt 0pt 90pt]{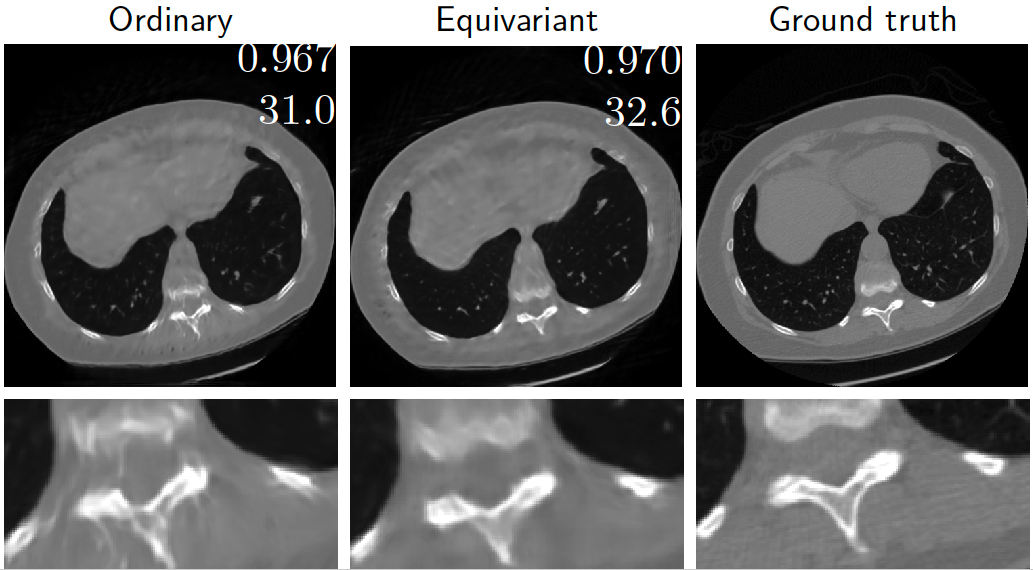}};%
    \node[rectangle split,rectangle split parts=2] at (-4.1,3.35) {\large Translational \nodepart{second} \large equivariance};
    \node[rectangle split,rectangle split parts=2] at (0,3.35) {\large Roto-Translational
  \nodepart{second} \large equivariance};
    \node at (4.1,3.15) {\large Reference};
    \end{tikzpicture}%
    \caption{Reconstruction with provable equivariant neural networks: translational versus roto-translational. Incorporating more inductive bias into the neural network improves the PSNR.\@ Even more importantly, the additional inductive bias leads to a better reconstruction of fine details. Data taken from~\cite{celledoni2021equivariant}.}
\label{fig:bydesign:ct}
\end{figure}
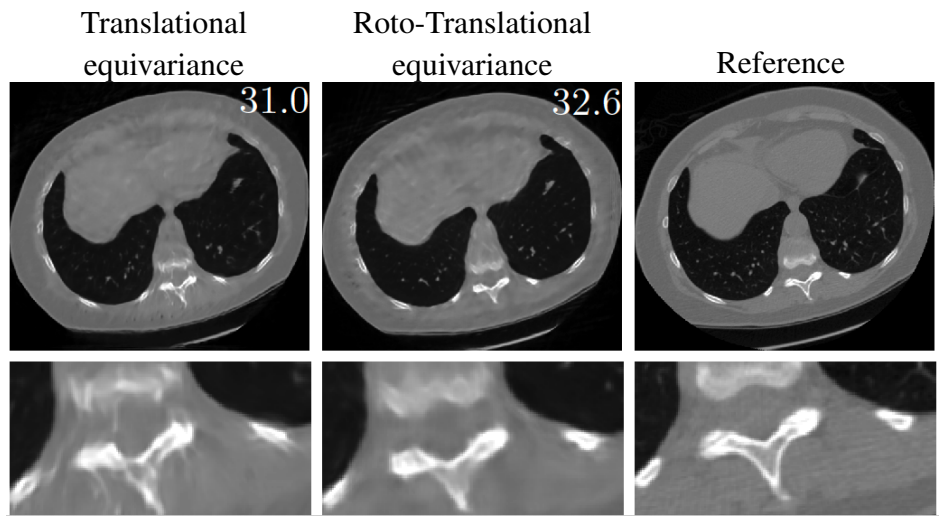

In the computational imaging setting, the presence of a non-equivariant forward operator can complicate the incorporation of natural symmetries into a reconstruction pipeline when thinking of the overall system, as highlighted in Figure~\ref{fig:equiv_summary}. Nevertheless, physics-driven network architectures such as those described in Section~\ref{sec: DL for CI} use components such as denoisers, or gradients of regularization functionals, which do naturally satisfy equivariance properties when the data has the corresponding symmetry. Indeed, let us take a statistical viewpoint: suppose we have a group $G$ acting on clean signals $u^*\sim p(u^*)$ and noisy signals $y | u^* \sim p(y|u^*)$ through $T$, and that these distributions are invariant, in the sense that $p(T_g(u^*)) = p(u^*)$ and $p(T_g (y) | T_g(u^*)) = p(y | u^*)$. If we measure the performance of a denoiser with an invariant loss function $\ell$, $\ell(T_g(u), T_g(u')) = \ell(u, u')$, the Bayes-optimal denoiser
\begin{align}
    \hat u_\text{Bayes} = \argmin_u \mathbb E_{(u^*, y)\sim p(u^*) p(y | u^*)} [\ell (u(y), u^*)]
\end{align}
will be equivariant:
\begin{align}
    T_g(\hat u_\text{Bayes}(y)) = \hat u_\text{Bayes}(T_g (y)).\
\end{align}
We can also motivate the use of equivariant operators from the perspective of the MBIR approach described in Section~\ref{sec: general CI}: if the regularization functional $J$ is invariant to a group symmetry, $J(u) = J(T_g(u))$, its proximal operator is equivariant~\cite{celledoni2021equivariant}. A hand-crafted regularization functional with such invariances is TV: it is invariant to translations, rotations and reflections, so that its proximal operator is equivariant with respect to these symmetries. Based on this property we are naturally led to use equivariant components in an unrolled learned iterative reconstruction algorithm, as was done in~\cite{celledoni2021equivariant}. Here proximal operators were modeled with roto-translationally equivariant components, as opposed to the usual CNN components which are just translationally equivariant, and it was observed that the additional symmetry allows for improved reconstruction quality (see Figure~\ref{fig:bydesign:ct}), greater data efficiency, and more robustness to images not seen in training.

\begin{figure}[ht!]
    \centering
    \begin{subfigure}{0.48\linewidth}
    \includegraphics[scale=1]{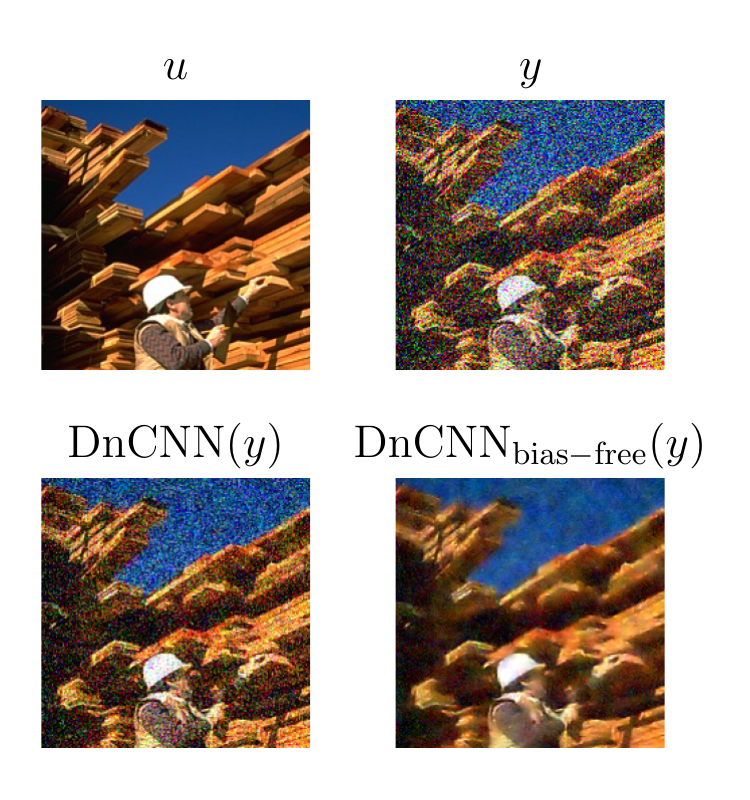}
    \end{subfigure}
    \begin{subfigure}{0.48\linewidth}
    \includegraphics[scale=1]{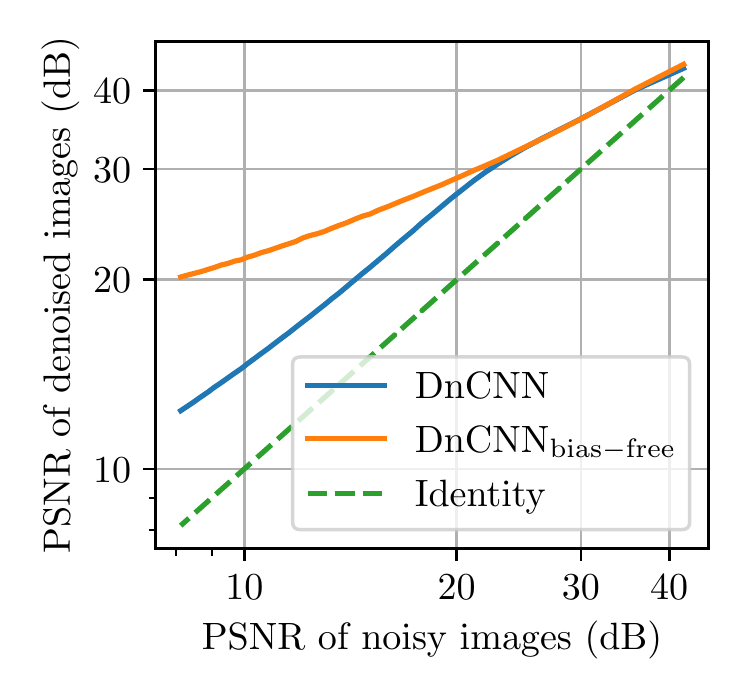}
    \end{subfigure}
    \caption{A comparison of DNN denoisers trained to denoise images corrupted by additive Gaussian white noise, using $\LeakyReLU$ as activation function. The denoisers are trained on pairs of clean and noisy images with a limited range of noise levels (PSNR $\sim 26-34$ dB) and then tested on a wide range of noise levels. Evidently, the denoiser that does not use biases (which is equivariant to scaling of the range) is vastly more robust to unseen noise levels than the denoiser that does use biases. \FS{The noisy image $y$ in this example has a PSNR of 9.9 dB, $\mathrm{DnCNN}(y)$ has a PSNR of 15.2 dB and $\mathrm{DnCNN}_\mathrm{bias-free}(y)$ has a PSNR of 21.0 dB.}}
\label{fig:bydesign:biasfree}
\end{figure}

As another interesting example of the use of equivariant DNN architectures in computational imaging, consider that many existing DNN architectures use the rectified linear unit, $\ReLU(x) = \max\{0, x\}$, or its ``leaky'' version, $\LeakyReLU_a(x) = \max\{ax, x\}$ for some $0<a<1$, as activation functions. These functions are positively 1-homogeneous, e.g.\ $\ReLU(cx) = c\ReLU(x)$ for $c>0$. In other words, if we let the group $G = (0,\infty)$ (with multiplication as the group product) be the group of scalings, acting on a vector $u\in \eR^d$ simply by scalar multiplication (or on a continuous signal by scalar multiplication on the range), $\ReLU$ and $\LeakyReLU$ (acting componentwise, as usual) are equivariant. In addition, any linear operator $A:\eR^d\to \eR^{d'}$ is trivially equivariant in this sense. Hence, ordinary DNN architectures are also equivariant in this sense, as long as no biases are used. This equivariance property has been exploited to design neural network denoisers that are robust to noise levels not seen in training~\cite{mohanRobustInterpretableBlind2020}, as shown in more detail in Figure~\ref{fig:bydesign:biasfree}. \MDcomment{It has also been used in GE Healthcare's CNN-based image enhancement algorithm, Air\textsuperscript{\texttrademark} Recon DL, for ringing suppression and SNR improvement that is embedded in their MR image reconstruction pipeline~\cite{Lebel2020}.}

\section{Equivariance by Learning}\label{sec: equiv learning}
An alternative way to impose equivariance is to enforce it through the training process instead of using equivariant architectures.
In the supervised setting where ground truth images are available, this can be done through DA. While for unsupervised learning a system-equivariant self-supervised loss can be used. The equivariance by learning methods presented in this section are architecture agnostic and can in principle be applied to a wide variety of machine learning models, \MDcomment{exploiting the acquisition physics within the associated training processes.}

\subsection{Equivariance Through Data Augmentation}

DA has a long history in machine learning as a way to expand the size of a limited dataset and is a research topic in its own right. It is based on the assumption that there is often additional information within the training data that has so far been unused. While the primary purpose of DA is to artificially increase the size of the dataset in order to improve the generalization properties of the network, it has strong links to notions of invariance/equivariance~\cite{Lenc2015}. Most applications of DA have focused on the task of classification (which we can associate with invariance rather than equivariance) however there is a growing interest in its use for image enhancement~\cite{Lebel2020} and in computational imaging, e.g.~\cite{Lee2018_mri,fabian2021MRAugment} where there is usually limited ground truth data.

The basic idea of DA is to introduce a set of transforms through which one can modify the existing training data to generate new plausible samples.
For example, as mentioned in the introduction AlexNet~\cite{krizhevsky2012imagenet}, increased the ImageNet data by a factor of 2048 through various transformations resulting in approximate equivariance~\cite{Lenc2015}. While modern DA extends to less well understood exotic transformations, including feature space and GAN-based augmentation, we will restrict our discussions here to those that can be directly linked to a group action and applicable in computational imaging. For a comprehensive review of other DA methods we refer the reader to \cite{shorten2019DAsurvey}.

\begin{figure}[t]
\centering
\includegraphics[width=0.6\textwidth]{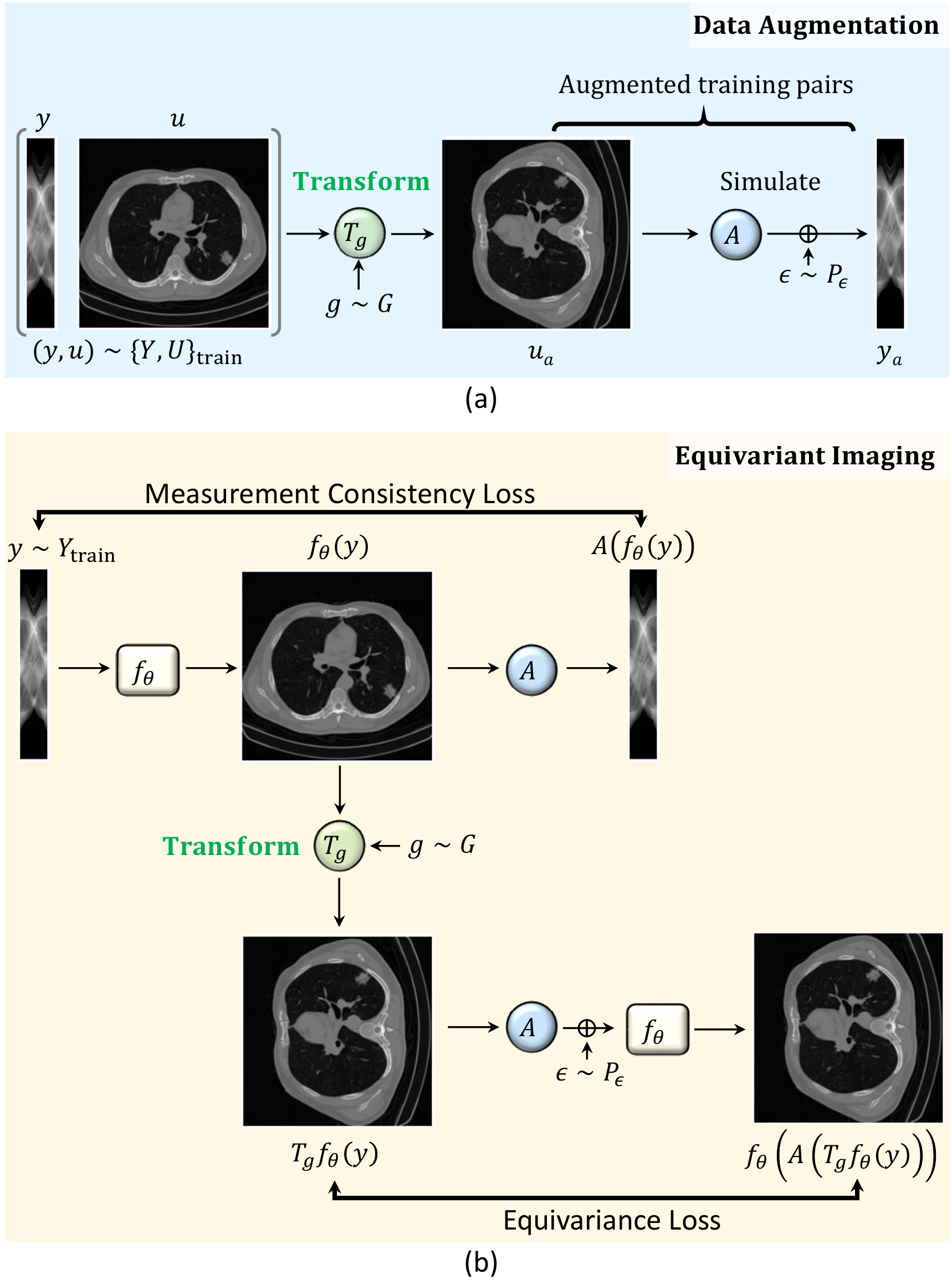}
\caption{A illustration of \emph{Equivariance by Learning}. (a) DA applied to training images and using simulation to synthesise augmented measurements; (b) Fully unsupervised EI strategy \cite{chen2021ei}. At the end of training, everything but $f_\theta$ is discarded, and $f_\theta(y)$ is used as the direct reconstruction function. In both (a) and (b), $\epsilon$ represents the measurement noise sampled from the distribution $P_\epsilon$.}
\label{fig:ei_diagram}
\end{figure}

Given a target image from training data for a classification task, one way to generate a new synthetic data sample is to apply an appropriate group action to the image. This transforms the input sample while the target output (the label) is expected to remain invariant. In computational imaging, the situation is significantly different. First of all, the image sample in this scenario serves as the target output with the associated raw measurements as the input to the reconstruction network.
Furthermore, unlike in the classification task, the associated measurements do not remain unchanged. It is therefore necessary to generate new measurements for the transformed image. This can typically be done through simulation exploiting the acquisition physics, including any noise process and system uncertainty. The core DA idea is illustrated in \Cref{fig:ei_diagram} (a).

\MDcomment{The benefits of DA for DNN-based accelerated MRI reconstruction were recently investigated in \cite{fabian2021MRAugment} using a state-of-the-art variational network \cite{Sriram2020}, modelling the complex valued imaging system and using a range of geometric transforms, including shearing and scaling, along with appropriate interpolation and anti-aliasing. The authors evaluated performance on highly undersampled ($8$-fold acceleration) single coil and multi-coil imaging and observed for training sets of up to $\sim 4K$ images, the inclusion of DA significantly improved performance and generalization, reducing the tendency of the network to overfit. Overall, the inclusion of DA achieved comparable structural similarity index measure (SSIM) performance to training without DA on a dataset $4\times$ - $10\times$ the size in both single coil and multi-coil experiments.}

An additional benefit of DA is that it provides a flexible way of building robustness into the learning process even beyond groups actions. \MDcomment{For example, GE’ Healthcare's image enhancement network within their  reconstruction pipeline~\cite{Lebel2020} used DA drawn from not just MR simulations with different rotations
and flips, etc., but also different pulse sequences, intensity gradients and contrast weightings. This can provide the network with important robustness to variations in the acquisition physics.}

\subsection{Equivariance in Unsupervised learning}\label{sec: equiv unsupervised learning}

\MDcomment{Being able to learn from measurement data alone is extremely desirable in various \corr{ill-posed} imaging problems} where obtaining ground-truth reconstructions to learn the signal model might be very expensive or even impossible. For example, this is often the case in medical and astronomical imaging.

\JT{Unfortunately, it is impossible to train a reconstruction network from only incomplete measurement data without any additional assumptions, even in the absence of noise~\cite{chen2021ei}
\corr{\footnote{\corr{In some circumstances unsupervised learning is possible when there is access to a range of different measurement operators, e.g. different k-space subsampling patterns in MRI. However, here we consider the more usual and most challenging problem where the data is collected through a single fixed measurement operator.}}}.
To observe this, consider a naive unsupervised loss which only enforces measurement consistency, e.g.,
\begin{align} \label{eq:naive unsup}
\sum_{i=1}^N \| Af_\theta(y_i)-y_i \|^2_2 .
\end{align}
This loss does not contain any information about the signals or their reconstructions in the nullspace of $A$, and thus there are infinitely many solutions $f_\theta$ which attain zero training error, including the trivial pseudo-inverse $f_\theta(y) = A^{\dagger}y$.}

Perhaps surprisingly though, the weak assumption of invariance to  actions \corr{of compact groups} can be enough for fully unsupervised learning~\cite{chen2021ei}. To understand this, note that such invariance means an observation $y$ can be equally thought of as an observation of a different signal, $\tilde{x}$, via a \emph{virtual} measurement operator $A_g = AT_g$ such that:
\begin{align}
    y = Ax = AT_g T_g^{-1} x = A_g\tilde{x}
\end{align}
where group invariance ensures that \corr{$\tilde{x} =T_g^{-1}x=T_g^Tx$} is a valid element of our signal model.
The group action here \emph{rotates} the nullspace of $A$:
\begin{align}
    \mathcal{N}_{A_g} = T_g^T \mathcal{N}_A
\end{align}
potentially exposing parts of the original nullspace to view. In order to see the whole of the signal space it is therefore necessary that the concatenation of all the virtual measurement operators,
\begin{equation} \label{eq:M matrix}
  M= \begin{bmatrix}
     AT_1 \\
     \vdots \\
     AT_{\ntransf}
    \end{bmatrix}
\end{equation}
be full rank. One can also think of $M$ here as being the combined measurement operator associated with having oracle simultaneous access to all the virtual measurements of the same signal, $x$.

A necessary condition for equivariant unsupervised learning is thus that $M\in \eR^{m\ntransf \times n}$ has a trivial nullspace, or in other words, that has rank $n$~\cite{chen2021ei}. This in turn implies that the group must be big enough such that $\ntransf>n/m$. However, not any combination of forward operator and group action verify this condition. Indeed, irrespective of the size of the group, if $A$ is itself equivariant then there exist actions $T_g'$ such that, $AT_g = T_g'A$, and we have $M = SA$ where $S\in\eR^{m\ntransf \times m}$ is the matrix containing $T_1'$ to $T_{\ntransf}'$. Thus $\rk{M}=\rk{A}=m<n$. For example, translation invariance cannot be used in an unsupervised manner to learn  from rank-deficient Fourier based measurement operators (which is the case in deblurring, super-resolution and accelerated MRI), as such an operator is equivariant to translations.

\subsubsection{Equivariant Imaging}
If the forward operator is \emph{not} equivariant and the group is big enough, then we can expect to be able to learn from only measurements (an in-depth analysis of the necessary and sufficient conditions for unsupervised learning can be found in~\cite{tachella2022sampling}). The equivariant imaging (EI) framework~\cite{chen2021ei} offers an elegant way of pursuing system equivariance through self-supervised learning, by using following surrogate loss function:

\begin{equation}\label{eqs:ten_loss}
       \argmin_\theta  \sum_{i=1}^N \sum_{g\in\group} \| Af_\theta(y_i) - y_i \|_2^2 + \alpha \| f_\theta(AT_gf_\theta(y_i)) - T_gf_\theta(y_i) \|_2^2,
 \end{equation}
where the first term enforces data consistency (c.f. \Cref{eq:naive unsup}), the second term enforces system equivariance, and $\alpha$ controls the strength of equivariance loss.  The training procedure is illustrated in~\Cref{fig:ei_diagram} (b). \JT{If the transformation $T_g$ is not an exact permutation of the pixel grid (e.g., if it is a rotation by 45 degrees), it can still be applied using bilinear interpolation and zero-filling.}  Crucially, the EI loss is fully unsupervised as it requires access to only measurement data, \MDcomment{works with undersampled  measurements}, \JT{and can be applied to any $f_\theta$, including unrolled networks}.

\begin{figure}[t]
\begin{center}
\includegraphics[width=0.95\textwidth]{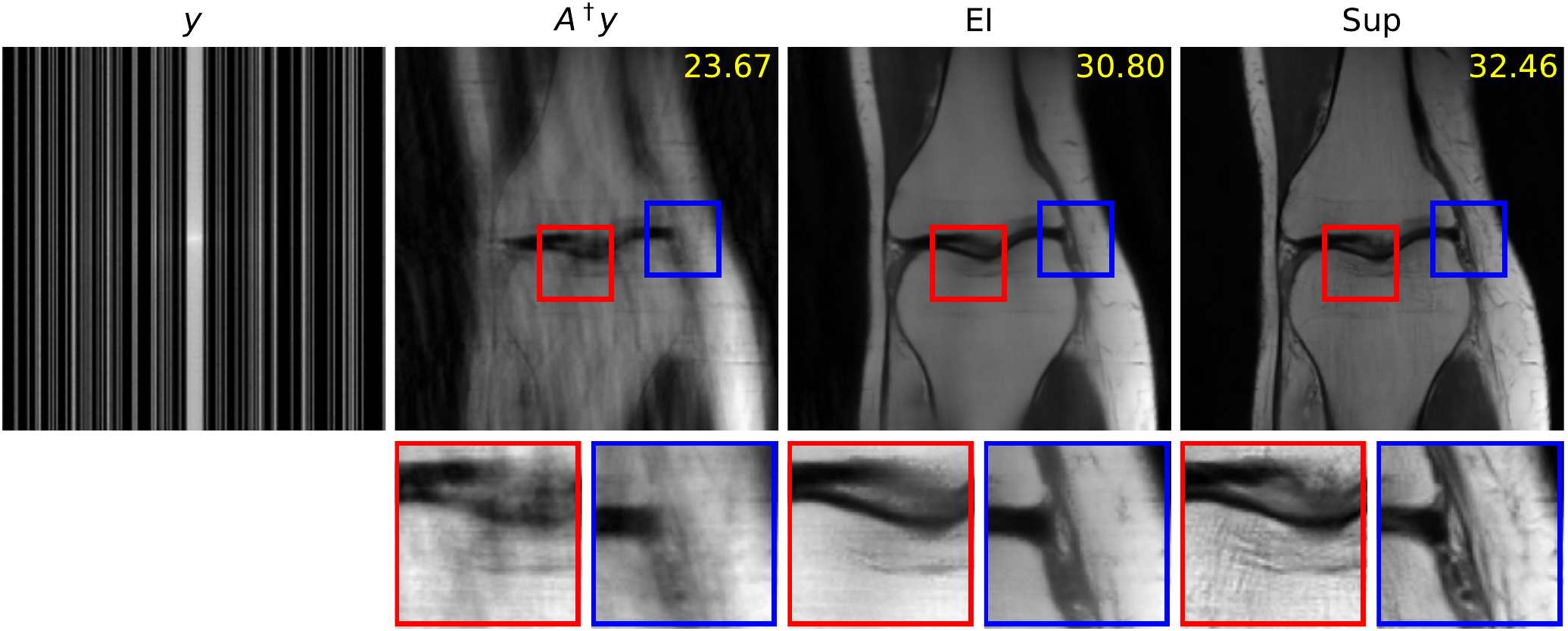}
\end{center}
\caption{A comparison of EI~\cite{chen2021ei} reconstruction with linear inversion and supervised learning for $4\times$ accelerated single coil MRI.
The network architecture in each case was an U-net taking $A^\dagger y$ as the input.
By enforcing system equivariance during unsupervised training, EI can perform almost as well as a fully supervised (Sup) network. PSNR values are shown in the top right corner of the images. 
} 
\label{fig: equiv imaging}
\end{figure}

\begin{figure}[t]
\centerline{\includegraphics[width=0.95\textwidth]{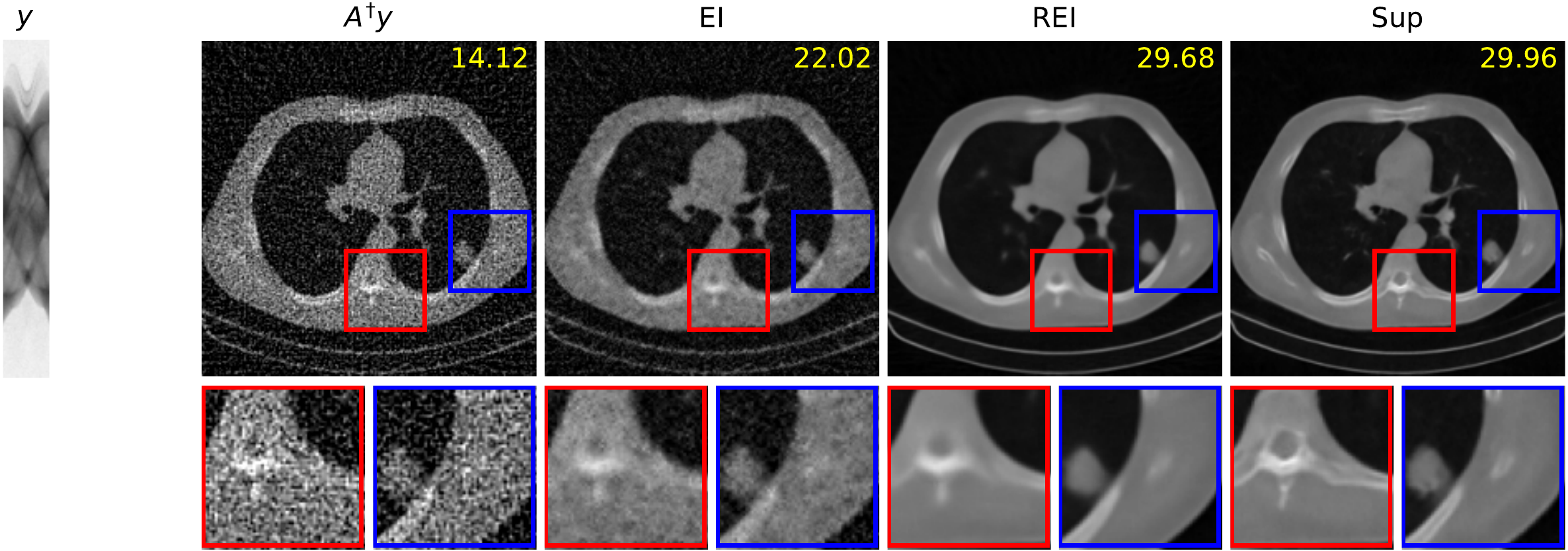}}
\caption{\emph{Low-dose} CT image reconstruction on the  test observations (50 views sinograms) with mixed Poisson-Gaussian noise. A comparison between linear inversion, EI, REI, and a supervised learning solution. PSNR values are shown in the top right corners. Data taken from~\cite{chen2021robust}.}
\label{fig:results_ct_rei}
\end{figure}

The EI approach was shown to achieve similar performance to supervised methods on a number of \MDcomment{underdetermined} image reconstruction tasks such as sparse-view CT, \corr{single coil} accelerated MRI (see \Cref{fig: equiv imaging}) and image inpainting~\cite{chen2021ei}.
This approach can also be adapted to account for noise (as long as the noise characteristics are known, e.g., \MDcomment{Poisson noise with known rate}). The modified robust EI (REI) approach~\cite{chen2021robust} combines Stein's unbiased risk estimator and the equivariance property for fully unsupervised learning from noisy measurement data  and, as with its noiseless counterpart empirically attains similar performance to fully supervised learning (see \Cref{fig:results_ct_rei}).

\subsubsection{Related Ideas}
The idea of learning from only partial measurements dates back to the concept of blind compressed sensing~\cite{gleichman2011bcs}, for the specific case of learning sparse dictionary models, where the authors showed that fully unsupervised learning is impossible without imposing strong constraints on the dictionary. Here, we have seen that the weak assumption of invariance to group actions is one way to provide such additional information and that it extends to the much richer models provided by DNNs.

Self supervision is also an important area of research in computational imaging in its own right and has been used to learn plug-and-play~\cite{liu2020rare} and unrolled~\cite{yaman2020self} network solutions for accelerated MRI reconstruction without fully sampled reference data. However, these methods require that the \corr{subsampled measurement operators vary across observations}~\cite{tachella2022sampling} or will be unable to learn signal information within the null space of the forward operator as set out above. The EI approach shows that the mild assumption of equivariance is in many cases enough to enable such fully unsupervised learning, effectively providing the multiple (virtual) operators for free. As most natural signals present certain invariances this approach offers an elegant way to learn the signal model without ground-truth data and with minimal additional assumptions.

\section{Open Problems and Future Directions}\label{sec: open problems}
Harnessing equivariance in the various ways outlined in this paper has shown great potential for imaging, especially in applications where large amounts of training data are expensive to obtain. Throughout the paper, we have seen how  equivariance improves generalization, provides robustness to noise level, and can enable fully unsupervised learning from measurement data alone.
While we have focused primarily on linear imaging systems, all the methods reviewed here can in principle be applied to more general non-linear imaging problems. This research field is rich in both challenges and opportunities, which we summarize next.

\paragraph{Opportunities and Limitations of Equivariance}
 There are still many possible design challenge to consider when developing equivariant neural network solutions for computational imaging. \MDcomment{As discussed in Section~\ref{sec:equiv_design}, there are different approaches to designing equivariant DNNs, and no standard implementation has yet emerged. Furthermore, current implementations tend to impose only limited symmetry, e.g. rotations of multiples of 90 degrees, and in computational imaging there is no guarantee that using an equivariant prox network will result in a full system equivariance. In contrast, it is straightforward to implement learned equivariance with a much wider class of transforms. \corr{The EI self-supervised loss is also quite general and could be incorporated into a wide range of learning strategies to directly achieve system equivariance, including supervised and semi-supervised learning, handling multiple forward operators, etc..} However, learnt equivariance acts only on the training data and may not be as robust as equivariance by design. It would therefore be interesting to explore whether a combination of the two methods might offer some advantage.}

Other important questions are whether we can always expect to achieve system equivariance and/or whether it is always desirable. \MDcomment{In many cases we may only be able to expect approximate symmetry in the data, e.g. when we consider rotations of arbitrary angle.} Attempting to enforce equivariance on non-invariant datasets may reduce imaging performance and we might instead wish to implement a restricted equivariance~\cite{weiler_general_2019}. Similarly, while local structures in images may be fully equivariant, at a global scale an image may have a preferred orientation. For example, a rotation of 180 degrees transforms a 6 into a 9 which clearly changes the information content. How important such aspects are for imaging is as yet unclear.

There are also many unanswered theoretical questions both in terms of generalization and identifiability. For example, it would be useful to quantify the robustness and generalization benefits of such solutions and to extend the recent identifiability results for unsupervised learning with equivariance~\cite{tachella2022sampling} to richer classes of group actions and operators.

\paragraph{General Group Actions}
An interesting challenge is to account for group actions beyond rigid transformations such as translations and rotations. For example, local approximate equivariant could be exploited to encode certain self-similarity of objects. This idea lies at the heart of Mallat's scattering transform~\cite{mallat2012group}, which gives an image representation that is stable (approximately invariant) to elastic deformations. In the context of computational imaging, we may ask instead for approximate equivariance to such deformations.
Alternatively, we could consider learning of the group actions themselves as part of the training. Recent research~\cite{keller2021topographic} has shown that this might be a feasible option.

\paragraph{Beyond Euclidean Domains}
The focus of the literature so far has been on scalar valued imaging, e.g., with applications such as MRI or CT.  Extensions to either the domain or range being a manifold or a graph are challenging \MDcomment{and fall within the emerging framework of geometric deep learning~\cite{Bronstein2021}. For example, the recently developed gauge CNNs~\cite{weiler2021} offer the possibility to build equivariant solutions on general Riemann manifolds. The ability to build more flexible equivariant networks could }
open up new opportunities for equivariant imaging \MDcomment{for challenging inverse problems such as  diffusion MRI or  point cloud data in lidar imaging.}

\section*{Acknowledgments}

MJE acknowledges support from the EPSRC (EP/S026045/1, EP/T026693/1, EP/V026259/1) and the Leverhulme Trust (ECF-2019-478). DC, MD and JT acknowledge support by the ERC C-SENSE project (ERCADG-2015-694888). MD is also supported by a Royal Society Wolfson Research Merit Award. FS acknowledges support from the EPSRC. CBS acknowledges support from the Philip Leverhulme Prize, the Royal Society Wolfson Fellowship, the EPSRC advanced career fellowship EP/V029428/1, EPSRC grants EP/S026045/1 and EP/T003553/1, EP/N014588/1, EP/T017961/1, the Wellcome Innovator Award RG98755, the European Union Horizon 2020 research and innovation programme under the Marie Skodowska-Curie grant agreement No. 777826 NoMADS, the Cantab Capital Institute for the Mathematics of Information and the Alan Turing Institute.

\bibliographystyle{IEEEtran}
\bibliography{IEEEabrv,reference}

\end{document}